\documentclass[aps,prl,twocolumn,groupedaddress,showpacs]{revtex4-1}

\usepackage{amsmath}
\usepackage{graphicx}
\usepackage{mwe}
\usepackage[caption=false]{subfig}
\usepackage[usenames, dvipsnames]{color}
\usepackage{verbatim}

\begin{document}

%\preprint{}

\title{Surface creasing of soft elastic continua as a Kosterlitz-Thouless transition}


\author{T. A. Engstrom}
\email[]{taengstr@syr.edu}
%\thanks{}
\author{J. M. Schwarz}
\email[]{jmschw02@syr.edu}
%\thanks{}
\affiliation{Soft Matter Program and Department of Physics, Syracuse University, Syracuse, NY 13244, USA}


\date{\today}

\begin{abstract}
Harnessing a model from composite materials science, we show how point-like cusped surface features arise as quasi-particle excitations, termed ``ghost fibers", on the surface of a homogeneous soft elastic material. These deformations appear above a critical compressive strain at which ghost fiber dipoles unbind, analogous to vortices in the Kosterlitz-Thouless transition. Finite-length creases can be described in the same framework. Our predictions for crease surface profiles and onset strain agree with previous experiments and simulations, and further experimental tests are proposed.\end{abstract}

\pacs{46.25.-y, 46.35.+z, 62.20.mq, 68.35.Rh, 82.60.Nh, 87.15.Zg} 
% elasticity in continuum mechanics of solids (first two) 
% buckling: structural failure of materials
% phase transitions at surfaces and interfaces 
% nucleation: chemical thermodynamics of
% phase transitions in biological systems
%\keywords{}

\maketitle


Cusped inward folds known as creases form on compressed surfaces of a variety of soft elastic materials \cite{jin14}, including natural rubber \cite{sou65,gen99}, polymer gels \cite{tan87,tru08,yoo10}, silicone elastomers \cite{cai12,che12,che14,lia16,tal14,tal16}, starchy foods \cite{hon09,cai10}, and the developing mammalian brain \cite{tal13,tal14,tal16,fer16}. In the latter context, creases are called ``sulci". Unlike the long-wavelength buckling of a compressed beam, or the smooth sinusoidal wrinkles observed on the skin of drying fruit or a tensioned elastic sheet \cite{cer03,hua07,pau16}, creases are sharply localized in both their elastic deformation and stresses, thereby defying a linear perturbation analysis \cite{gen99,hon09,hoh11,hoh12,cai12}. Owing to this difficulty, numerical minimization of a nonlinear neo-Hookean energy functional has become the standard theoretical tool for investigating the onset of creases \cite{hon09,ben10,hoh11,hoh12,cai12,tal13,tal14,tal16}. A central claim in much of this work is that creasing is a fundamentally new, nonlinear instability with no scale \cite{hoh11,hoh12}. Experimental work has also studied the growth of pre-existing long creases, describing these behaviors in analogy to crack propagation \cite{che12,che14}.

Here we develop a new quasi-particle framework for shear stress focusing in surface-compressed solids, assuming planar geometry and neglecting surface tension. We apply our theory to the  creasing instability, obtaining a markedly different picture than \cite{hoh11,hoh12}. We find evidence that (i) creasing onset maps to the Kosterlitz-Thouless (KT) transition \cite{kos73}, (ii) nonlinear elasticity is needed only within a small region analogous to a vortex core, and (iii) compression-induced shear strain fluctuations set the fundamental, microscopic lengthscale in the problem. Our theory makes contact with experimental and simulation results on critical strain, surface profiles, and crease patterns.  In particular, we obtain a universal critical compressive plane strain above which creases emerge, in reasonable agreement with the measured value of $35\%$ \cite{tru08,hon09,hoh12}. Finally, the theory points to a set of minimal physical ingredients for creasing, and suggests a possible unification with ridging (formation of localized surface protrusions) \cite{tak14}, and dimple crystallization \cite{bro15,jim16}.

\begin{figure}[b]
\centering
\includegraphics[width=0.48\textwidth]{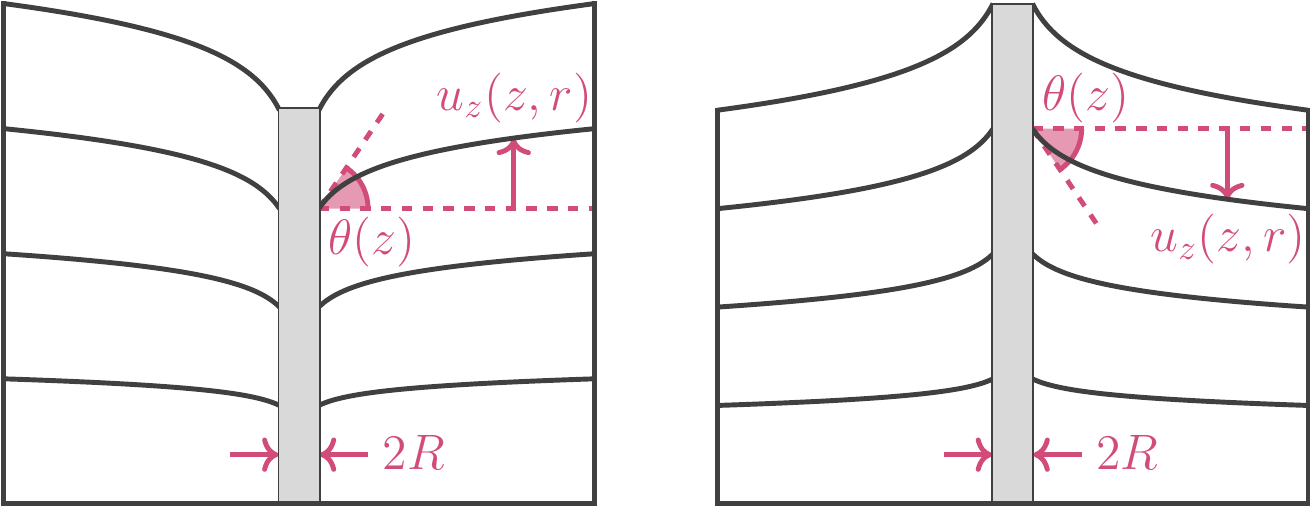}
\caption{\label{cartoon}
Cartoon of axisymmetric shear lag. Curved lines indicate the matrix deformation $u_z\sim\tan(\theta)\ln(r/R)$ around an isolated vertical fiber of radius $R$ (shaded). The left case shows an innie deformation that results when the fiber is under axial tension; the right case shows an outie resulting from a fiber under compression.}
\end{figure}

Our point of departure from prior work is to consider a distinct regime of zero-length creases, qualitatively similar to those observed in \cite{yoo10,cai12,che12}, immediately upon nucleation, and those in \cite{tal13}, as the critical point is approached from above. Foundational to our theory is the observation that zero-length creases and zero-length anticreases (hereafter, innies and outies) also appear in a very different continuum elastic context, namely the shear lag model of composite materials science and engineering \cite{cox52, hul96}. In this model, one assumes that shear coupling is supported at the interface between a low-dimensional reinforcing phase (i.e., 1d fibers or 2d slabs) and a surrounding 3d matrix phase. Next, an approximation is made that the transfer of axial loads between the two components is accomplished entirely via tension or compression in the reinforcing phase, and pure shear in the matrix. Axial loads refer to external or internal forces (such as those arising from differential growth of the two components) acting parallel to a long axis of the reinforcing phase. In the case of a fiber-matrix composite, the model predicts that matrix shear stress and strain fall off as $1/r$, where $r$ is the perpendicular distance from a fiber. Thus, the matrix deformation and hence the surface profile scales as $\ln r$ (see Figure \ref{cartoon}).   

%%%
\paragraph*{Mapping shear lag to 2d electrodynamics ---}
%%%

Let us take all forces in the shear lag model along $z$. In the matrix phase, the only non-negligible components of the strain tensor $\epsilon_{ij}=(\partial_iu_j + \partial_ju_i)/2$ have one index equal to $z$ and the other not equal to $z$. Defining a 2d vector of shear strains $\vec{\gamma}=2(\epsilon_{xz},\epsilon_{yz})$, force balance on a volume element of matrix takes the form of Gauss's Law
\begin{equation}
\oint \vec{\gamma}\cdot d\mathbf{A} = \frac{f_0}{G}.
\end{equation}
Here $G$ is the matrix shear modulus while $f_0$ is the net force supported at the interface and enclosed by the free body diagram (Gaussian surface).  In terms of the wetted perimeter $p_0(z)$ and interfacial shear stress $\tau_0(z)$, the interfacial force is $f_0=\int dz\,p_0\tau_0$. Our neglect of $\epsilon_{zz}$ is justified by requiring $p_0\tau_0$ vary slowly with $z$.

Equivalently, one has a Poisson's equation for the (scalar) deformation field $u_z$. For the important special case of a thin fiber source at the origin, 
\begin{equation}
\nabla^2_ru_z(z,\mathbf{r}) = 2\pi R\tan(\theta(z))\delta^2(\mathbf{r}), \label{poisson}
\end{equation}
where $R$ is the fiber radius and $\tan\theta=\tau_0/G$ is the interfacial shear strain, as indicated in Figure \ref{cartoon}. $2\pi R\tan\theta$ is the ``charge" per unit length of fiber. (Note that in the case of a viscoelastic matrix with storage and loss moduli given by $G'$ and $G''$, respectively, a frequency-dependent ``dielectric function'' appears as $1+iG''/G'$.) The solution of Equation \ref{poisson} is
\begin{equation}
u_z(z,\mathbf{r}) = -2\pi R\tan(\theta(z))C(\mathbf{r}),
\end{equation}
where $C(\mathbf{r}) = -\ln(|\mathbf{r}|/R)/2\pi$ is the 2d Coulomb potential (Green's function).

Guided by this mapping, we ask whether $u_z$ might be associated with a quasi-charge excitation in which the fiber of the conventional shear lag model is an abstraction. For system size $L$ and vertical thickness $h$, the elastic strain energy (electrostatic energy) $\frac{G}{2}\int dV\,\gamma^2$ required to create an isolated such ``ghost fiber" is
\begin{equation}
U_{\textrm{gf}} = E_{\textrm{core}} + \pi GhR^2\,\overline{\tan^2\theta}\ln(L/R).\label{energy}
\end{equation}
The overbar denotes an average over $z$, and the core contribution is from deformations within $r<R$. Notably, this result resembles the energy of an isolated vortex \cite{kar07}, and $\frac{G}{2}\gamma^2$ is the leading order term of the energy density used in the aforementioned neo-Hookean simulations imposing an incompressibility constraint. Motivated thus, we now turn to the statistics of ghost fibers.

%%%
\paragraph*{Effective thermodynamics of shear lag quasiparticles  ---}
%%%

Creasing in a flat geometry requires some spatial inhomogeneity to break translation invariance and nucleate the creases. Typically there is disorder at both the mesoscale and microscale, due to dust, surface imperfections, and the polymer network itself. In experiments, nucleation sites have also been introduced in a controlled manner via surface-embedded microspheres \cite{yoo10}. Creasing simulations rely upon defect seeding \cite{cai12}, inhomogeneous surface normal forces \cite{hoh12}, random vertical displacements \cite{tal13}, or random variations in growth \cite{tal14}. In all such cases, these defects pattern the surface with stress concentrators that give rise to strain fluctuations within a thin subsurface layer, when the system is macroscopically compressed.

The central postulate of this work is that disorder and drive-induced shear strain fluctuations (from one or more sources) can play the role of thermal fluctuations, i.e. there is an ``effective temperature'' that increases during a slow compression of a disordered elastomer or gel, enabling a strongly \textit{athermal} system of ghost fibers to access many microstates (position and charge configurations). Similar approaches have been used to describe the statistics of granular materials, pinned vortex lattices, artificial spin ice, and other athermal systems \cite{bi15,cug11,cas03,nis10,kol02}. Here, our effective temperature postulate is of the form 
\begin{equation}
k_BT_{\textrm{eff}} \sim G \langle\epsilon_d^2\rangle \lambda_d^2\,l_d, \label{Teff}
\end{equation}
where $\langle\epsilon_d^2\rangle$ is the mean squared amplitude of $\epsilon_{xz}$ and $\epsilon_{yz}$ fluctuations, and $\lambda_d$, $l_d$ are the characteristic wavelength and skin depth of these fluctuations. Equation \ref{Teff} is reminiscent  of the Lindemann criterion for bulk melting of a harmonic solid \cite{lin10}. In the following, we will identify the effective system thickness $h$ with $l_d$, below which the system is cold and inactive.

Working within the microcanonical ensemble, the configurational entropy of a system containing a single ghost fiber is $S = 2k_B\ln (L/R)$. The free energy cost to create the ghost fiber is $F = U_{\textrm{gf}} - T_{\textrm{eff}}S$.  In the thermodynamic limit where the finite core energy is dominated by the logarithmically divergent term, $F<0$ for mean square strain fluctuations greater than the critical value
\begin{equation}
\langle\epsilon_d^2\rangle_c = \frac{\pi}{2}\frac{R^2}{\lambda_d^2}\,\overline{\tan^2\theta}. \label{criticalstrain}
\end{equation}
Because the quantity that maps to electric charge is an odd function of $\theta$, a charge dipole corresponds to an innie-outie pair of surface deformations, equivalent to a tension-compression pair of ghost fibers (see Figure \ref{cartoon}).  The energy required to create a pure dipole is finite, in contrast to Equation \ref{energy}, and thus 3d ghost fibers are analogous to 2d vortices: for $\langle\epsilon_d^2\rangle < \langle\epsilon_d^2\rangle_c$, the system contains tension-compression bound pairs of ghost fibers, and at $\langle\epsilon_d^2\rangle_c$ there is an unbinding transition (KT transition).

We now consider the grand canonical ensemble. A charge neutral system of ghost fibers has Hamiltonian
\begin{equation}
H = \sum_iE_{\textrm{core},i} + 4\pi^2Gh\sum_{i<j} R_iR_j\overline{\tan\theta_i}\,\overline{\tan\theta_j} C(\mathbf{r}_i-\mathbf{r}_j).
\end{equation}
In the regime where elastic deformations within the core are linear, or weakly nonlinear, or there are strong nonlinearities but they are confined to $r\ll R$, simple scaling arguments indicate $E_{\textrm{core}} \sim GV_{\textrm{core}}\epsilon_{\textrm{core}}^2 \sim Gh\pi R^2(\overline{\tan\theta})^2$. This quantity would appear to vary from one quasi-particle to another because the charges $\sim R_i\overline{\tan\theta_i}$ are here continuous degrees of freedom. However, we can exploit the arbitrariness of the $R_i$ in order to take the core energy as a meaningful chemical potential $\mu$. The appropriate choice is $R_i=s|\overline{\tan\theta_i}|^{-1}$, where $s\sim\sqrt{\mu/(Gh\pi)}$. This brings the partition function into the Coulomb gas form
\begin{equation}
Z = \sum_{\{n_i\}} \int \prod_i d^2r_i \;y_0^{\sum_i n_i^2} e^{4\pi \ln y_0 \sum_{i<j} n_in_j C(\mathbf{r}_i-\mathbf{r}_j)}, \label{partitionfct}
\end{equation}
where $n_i=\pm1$ and $y_0=\exp[-\mu/k_BT_{\textrm{eff}}]$ is the ghost fiber fugacity. The price paid for replacing continuous charges with discrete ones is that $R$ is now an ambiguous ``lattice constant". However, we have made available a small and well-defined lengthscale $s$; this can presumably replace $R$ as the short distance cutoff. 

The fugacity $y_0$ and ``coupling constant" $K=-(\ln y_0)/\pi$ are related because we are considering a specific physical system (e.g. \cite{kar07}). Intersection of the line $y_0=e^{-\pi K}$ and the line of fixed points $y_0=-\pi^{-2}(K^{-1}-\pi/2)$ determines the critical inverse coupling $K_c^{-1}=1.06$, and hence the critical mean square strain fluctuation $\langle\epsilon_d^2\rangle_c=(s^2/\lambda_d^2)K_c^{-1}$. Note $K_c^{-1}$ is depressed from the mean field value $\pi/2$ obtained earlier.

Below $\langle\epsilon_d^2\rangle_c$, large-scale surface deformation would not be seen because the tension-compression pairs are tightly bound. The appearance of spatially separated, cusped surface deformations at a critical point that has no explicit dependence on system thickness or shear modulus is consistent with creasing experiments \cite{tru08}. So too, we argue from data in \cite{cai12,che12,yoo10,tal13}, is the notion of zero-length creases at the critical point. (But see \cite{che14} for a different interpretation of crease lengths.) In light of the apparently universal 35\% onset strain \cite{tru08,hon09,hoh12}, one is tempted to identify $\lambda_d$ with the lattice constant $s$ (possibly scaled by a numerical prefactor). In fact, the simulations of Tallinen, et al. use random vertical displacements of mesh surface nodes, consistent with this picture \cite{tal13}. Their method of introducing fluctuations suggests a way to estimate the critical compressive plane strain $\epsilon_c$, given $\langle\epsilon_d^2\rangle_c$, but first we need a brief digression.

A consequence of all the stresses in the problem being shear stresses is that the area of the free surface is approximately conserved (within linear elasticity). One can see this by considering a 2d finite element in simple shear, and noting that its perimeter change is a second order effect. Here and in one place to follow, we shall make use of this approximate ``area conservation principle".

Returning to the problem at hand, we approximate the shear strain fluctuations by a square wave with amplitude $\sqrt{\langle\epsilon_d^2\rangle}$, such that the corresponding fluctuations in vertical displacement are a triangular wave with amplitude $\frac{1}{2}\lambda_d\sqrt{\langle\epsilon_d^2\rangle}$. Setting $s=\lambda_d/2$, as suggested by the method in \cite{tal13}, one then finds $\epsilon_c=1-(1+K_c^{-1})^{-1/2}=38\%$ (mean field theory) and $=30\%$ (renormalization group), which bracket 35\%.

%%%
\paragraph*{Post-KT dynamics of ghost fibers  ---}
%%%

Two features of creasing experiments and simulations remain to be explained by our quasi-particle theory: (i) that only innies and not outies appear to be seen, and (ii) that innies smoothly become finite-length creases. In this section we consider (i), and in the next section we will consider (ii).

The KT transition does not involve (or at least, does not require) self-contact in the core region. Yet self-contact is generically observed \cite{hon09,hoh12,che12,cai12,tal13}. We propose that self-contact ensues at strain $\epsilon_{sc}> \epsilon_c$, and point out that it can only be available to innies, because a self-contacting outie is an unphysical concept. Appealing to area conservation (now only qualitatively useful since the core deformations are strongly nonlinear), outies should incur a higher energy penalty than innies, in the regime $\epsilon>\epsilon_{sc}$,  because they are not as effective at sequestering surface area. The system cannot exactly get rid of its outies, however. Doing so would generate a nonzero net charge, causing $u_z$ to grow with system size as $(\sum_in_i)\ln L$, clearly inconsistent with creasing experiments. What happens, we propose, is that an outie's $R$ increases while its $|\overline{\tan\theta}|$ decreases, in such a way that its charge $-2\pi R|\overline{\tan\theta}|$ stays fixed. In other words, the negative point charges get smeared out into a negative background charge (cf. the non-neutral Coulomb gas \cite{min87}). The Poisson equation for this situation reads
\begin{equation}
\nabla^2_ru_z = 2\pi s \sum_{\textrm{innies}} \delta^2(\mathbf{r-r}_i)  -  \alpha,
\end{equation}
where $-\alpha$ is a uniform negative charge density, interpreted as the surface curvature the system would have, if the innies were removed. 

One might ask if this innie-outie shape asymmetry could be present even during the KT transition. We suggest the answer is no, because the ghost fiber bound pairs that exist below $\epsilon_c$ must have an essentially net zero surface deformation in order to be consistent with the observed flat surface. The shape asymmetry is something that arises in connection with the energy penalty for unable-to-self-contact outies. One might also ask if there exist conditions in which a system of \textit{outies} in a neutralizing \textit{positive} background is realized. Tentative support for this idea comes from recently observed ridges \cite{tak14}, which are somewhat reminiscent of anticreases.

%%%
\paragraph*{Ghost slabs  ---}
%%%

\begin{figure}[t]
\centering
\includegraphics[width=0.48\textwidth]{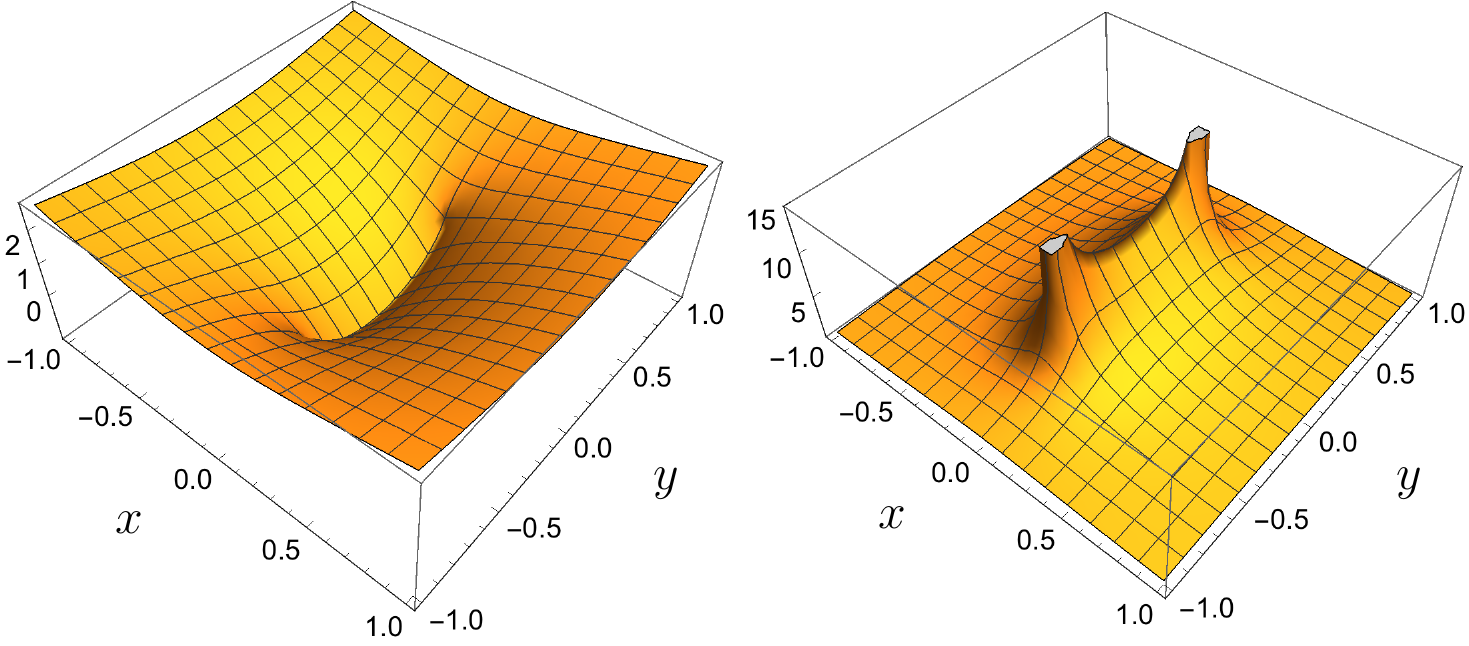}
\caption{\label{slab}Deformation field $u_z$ (left panel) and energy density $(\nabla u_z)^2$ (right panel) associated with a finite length ghost slab, for $\alpha=0$. Arbitrary units are used for the vertical axes while the horizontal axes are in units of $\ell$. The two peaks in the energy density are cut off for visualization purposes, however they are not singularities.}
\end{figure}

\begin{figure}[t]
\centering
\includegraphics[width=0.48\textwidth]{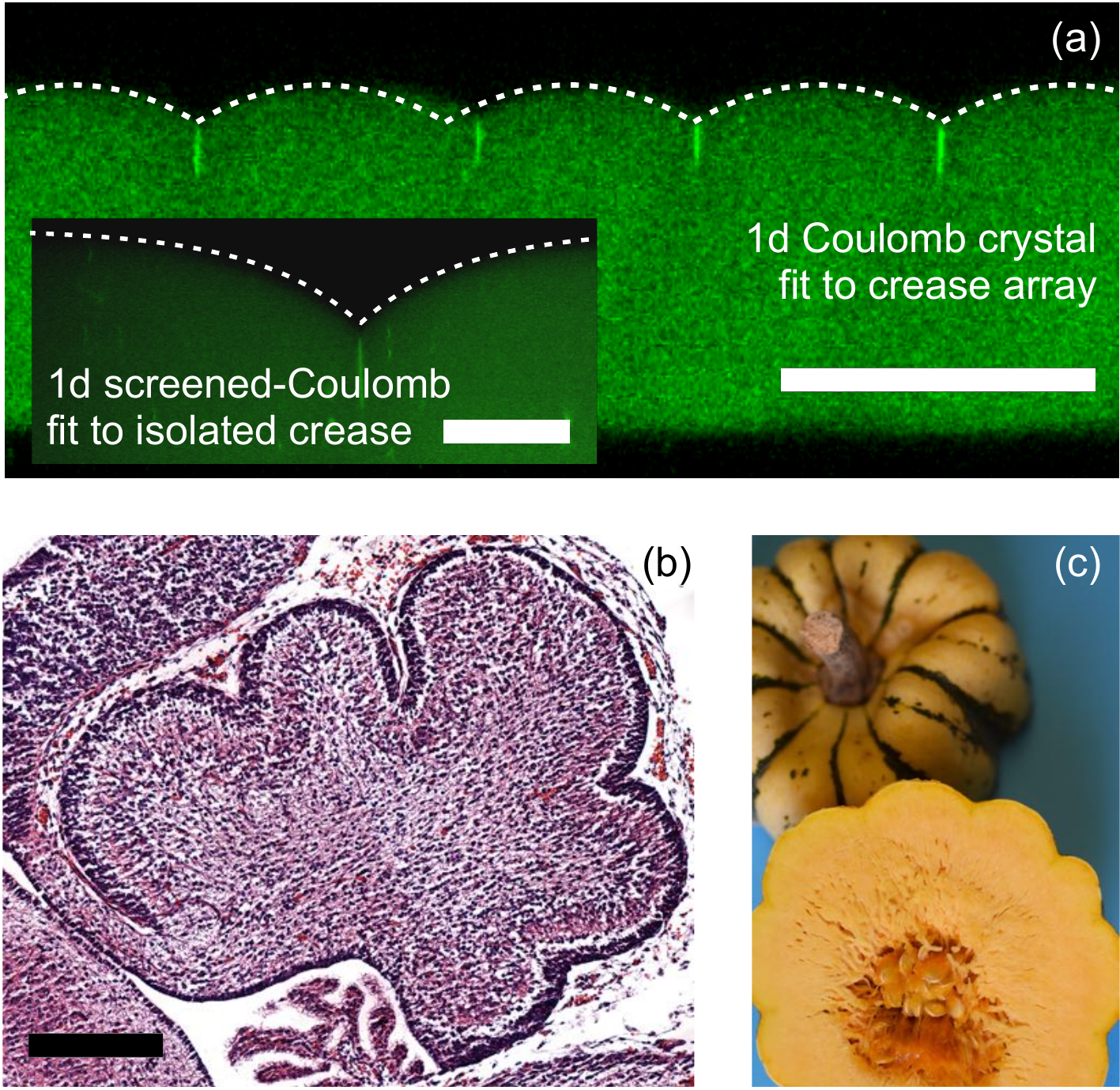}
\caption{\label{experiments}
(a) Confocal microscope image of a 1d crease array in a PDMS elastomer film under 55\% uniaxial compression (via attachment to a pre-stretched substrate). Inset: isolated crease in a PDMS film under 46\% compression. (Images courtesy of Dayong Chen and Ryan Hayward). In both images the crease(s) run perpendicular to the plane of the page, the small bright regions are regions of self-contact, and the scale bar indicates 40 microns. The 1d Coulomb crystal fit is of the form $u_z(x_i)\sim x_i-x_i^2/a$, as discussed in the main text, where $a$ is the average crease spacing taken from the experimental image, and $x_i$ is the spatial coordinate within the $i^{th}$ unit cell. (b) Midline saggital section of a mouse cerebellum at 18.5 embryonic days development, roughly 1-2 days after creasing onset (courtesy of Andrew Lawton and Alex Joyner). The scale bar indicates 200 microns. (c) Creasing in a sweet dumpling squash (courtesy of Indian Creek Farm, Ithaca, NY), elsewhere described as wrinkling \cite{yin08,wan15}.}
\end{figure}

Having considered ghost fibers, we now consider a thin ``ghost slab" of length $2\ell$ and height $h$ characterized by charge density $\rho\sim\delta(x)[H(y+\ell)-H(y-\ell)]$. (The case $\alpha\neq0$ will be treated momentarily). A straightforward application of Green's method yields material deformation
\begin{eqnarray}
u_z(\mathbf{r}) &\sim& (y+\ell)\ln\Big[(y+\ell)^2+x^2\Big] + 2x\tan^{-1}\Big(\frac{y+\ell}{x}\Big) \nonumber \\
 &-&(y-\ell)\ln\Big[(y-\ell)^2+x^2\Big] - 2x\tan^{-1}\Big(\frac{y-\ell}{x}\Big), \nonumber \\
 &+& \textrm{constant,} \label{slabshape}
\end{eqnarray}
and strain energy density
\begin{eqnarray}
(\nabla u_z)^2 &\sim& \Big[ \tan^{-1}\Big(\frac{y+\ell}{x}\Big) -  \tan^{-1}\Big(\frac{y-\ell}{x}\Big) \Big]^2 \nonumber \\
&+& \Big[ \textrm{tanh}^{-1}\Big(\frac{2\ell y}{r^2+\ell^2}\Big) \Big]^2. \label{slabenergydensity} 
\end{eqnarray}
Equations \ref{slabshape} and \ref{slabenergydensity} are plotted in Figure \ref{slab} for $\ell=1/2$. In the thermodynamic limit, the rotational contribution to the ghost slab entropy is insignificant, and the only important contribution to the elastic energy comes from the monopole term. Similar arguments to those used above lead to a critical strain fluctuation for slabs $\langle\epsilon_d^2\rangle_{c,\textrm{ slab}} \sim (\ell^2/\lambda_d^2) \,\overline{\tan^2\theta}$.  This result reveals ghost fibers to be a limiting case of ghost slabs (i.e., as $\ell\to R$).

Experiments and simulations employing uniaxial strain tend to generate straight, parallel creases; when the strain is large, the creases can span the system size, forming a 1d array \cite{cai12,che12,tal13}. Within our electrostatics framework, for $\alpha\neq0$, surface profiles of such systems are predicted to have the same form as the electrostatic potential of a 1d Coulomb crystal (a periodic stack of infinite, charged sheets embedded in a charge-neutralizing, uniform background). Figure \ref{experiments}a shows a test of this prediction, with no adjustable fit parameters apart from an overall prefactor. The possibility of the background being ``polarizable" is investigated via the $d$-dimensional charge screening equation $(\nabla^2_d-\lambda^{-2})\phi = -4\pi Q\delta^{(d)}$, where $Q$ is charge and $\lambda$ is the screening length \cite{joh12}. The $d$=1 solution, $\phi(x)\sim e^{-x/\lambda}$, is fit to the profile of an isolated long crease in Figure \ref{experiments}a inset. The qualitative shape of the crease array in Figure \ref{experiments}a (parabolic crests between sharp cusps) occurs in other settings such as a mouse cerebellum  and a winter squash (Figure \ref{experiments}b,c). Prior work has modeled these as elastic materials \cite{lej16,hu11}, suggesting the same mechanism may be at play, in spite of the different (curved versus planar) geometry.

Experiments and simulations employing equibiaxial strain tend to generate a square lattice of short, straight creases with each nearest neighbor pair having relative orientation of 90 degrees; at high strains, a hexagonal lattice of 3-fold symmetric, Y-shaped creases is also seen \cite{tal13}. (Figure 3c in \cite{tru08} exhibits both motifs.) Within our framework, such patterns again have a natural interpretation as charge-crystallized shear lag quasi-particles. In the limit where the spatial extent of quasi-particles is very small compared with their lattice spacing (i.e., they are ghost fiber-like), the creasing pattern is predicted to be a 2d Coulomb crystal with hexagonal symmetry (e.g. \cite{bon08}). In fact the hexagonal dimple crystal observed in \cite{bro15,jim16} is suggestive of this, although its discoverers give a wrinkling, not creasing, interpretation. Outside the limit of fiber-like quasiparticles, the ground state crystal structure is an interesting topic for future work. 

%%%
\paragraph*{Discussion  ---} 
%%%

Upon building a composite materials-inspired quasi-particle framework, we have uncovered a creasing scenario that involves at least three distinct regimes. For in-plane compressive strain $\epsilon<\epsilon_c$, the system contains tightly bound pairs of ghost fibers whose deformation fields largely cancel. For $\epsilon_c<\epsilon<\epsilon_{sc}$, the pairs are unbound, giving rise to spatially separated, cusped surface deformations. Also in this regime, innies (outies) smoothly become finite length creases (anticreases), as ghost fibers smoothly change dimensionality into ghost slabs. For $\epsilon>\epsilon_{sc}$, anticreases smear out into a charge-compensating background, while repulsive interactions between creases causes them to organize into a Coulomb crystal. The KT transition is predicted to occur at a universal critical strain $\epsilon_c=30\%$ (renormalization group analysis) and self-contact is expected to commence at a slightly higher strain $\epsilon_{sc}$, which we speculate may be the previously measured onset strain, $35$\%. Both transition points should be observable, by virtue of the characteristics of the regimes they delineate (e.g. surface profiles), providing a means for experimental tests of our theory.

\begin{acknowledgments}
The authors are grateful to Ryan Hayward and Joseph Paulsen for many helpful discussions, and acknowledge financial support from NSF-DMR-CMMT Award Number 1507938. 
\end{acknowledgments}

%\bibliography{creasingKT}
\providecommand{\noopsort}[1]{}\providecommand{\singleletter}[1]{#1}%

\end{document}